\documentclass[conference]{IEEEtran}
\IEEEoverridecommandlockouts
\usepackage{cite}
\usepackage{amsmath,amssymb,amsfonts}
\usepackage{algorithmic}
\usepackage{graphicx}
\usepackage{textcomp}
\usepackage{xcolor}
\def\BibTeX{{\rm B\kern-.05em{\sc i\kern-.025em b}\kern-.08em
    T\kern-.1667em\lower.7ex\hbox{E}\kern-.125emX}}
\begin{document}

\title{A Novel Stochastic Transformer-based Approach for Post-Traumatic Stress Disorder Detection using Audio Recording of Clinical Interviews\\
}

\author{\IEEEauthorblockN{1\textsuperscript{st} Mamadou Dia}
\IEEEauthorblockA{\textit{LISSI} \\
\textit{Université Paris-Est Créteil (UPEC)}\\
Vitry-sur-Seine, France \\
mamadou.dia@u-pec.fr}
\and
\IEEEauthorblockN{2\textsuperscript{nd} Ghazaleh Khodabandelou}
\IEEEauthorblockA{\textit{LISSI} \\
\textit{Université Paris-Est Créteil (UPEC)}\\
Vitry-sur-Seine, France \\
ghazaleh.khodabandelou@u-pec.fr}
\and
\IEEEauthorblockN{3\textsuperscript{rd} Alice Othmani}
\IEEEauthorblockA{\textit{LISSI} \\
\textit{Université Paris-Est Créteil (UPEC)}\\
Vitry-sur-Seine, France \\
alice.othmani@u-pec.fr}
\and
}

\maketitle

\begin{abstract}
Post-traumatic stress disorder (PTSD) is a mental disorder that can be developed after witnessing or experiencing extremely traumatic events. PTSD can affect anyone, regardless of ethnicity, or culture. An estimated one in every eleven people will experience PTSD during their lifetime. The Clinician-Administered PTSD Scale (CAPS) and the PTSD Check List for Civilians (PCL-C) interviews are gold standards in the diagnosis of PTSD. These questionnaires can be fooled by the subject's responses. This work proposes a deep learning-based approach that achieves state-of-the-art performances for PTSD detection using audio recordings during clinical interviews. Our approach is based on MFCC low-level features extracted from audio recordings of clinical interviews, followed by deep high-level learning using a Stochastic Transformer.
Our proposed approach achieves state-of-the-art performances with an RMSE of 2.92 on the eDAIC dataset thanks to the stochastic depth, stochastic deep learning layers, and stochastic activation function.
\end{abstract}


\section{Introduction}
\subsection{Post-Traumatic Stress Disorder}
\label{sec:PTSD}
Post-traumatic stress disorder (PTSD) is a condition that can arise after exposure to very traumatic experiences such as interpersonal violence, combat, life-threatening accidents, or a natural disaster.
Even after the horrific incident has passed, people with PTSD continue to endure intense, unsettling thoughts and sensations related to their experience. Flashbacks or dreams may cause them to relive the incident, they may experience sadness, fear, or rage, and they may feel distant or estranged from other people.

Assuming that exposure to trauma was the primary cause of PTSD,  initial studies \cite{yehuda200719} looked at the processes connected to stress reactions and traumatic memories.
All people, regardless of race, nation, or culture, can experience PTSD. An estimated one in 11 people will receive a PTSD diagnosis in their lifetime, and it affects about 3.5 percent of adult U.S. citizens each year \cite{ptsd2022Taylor}.

During a structured clinical interview, patients' self-report questionnaires are mostly used to make the diagnosis of PTSD. The Clinician-Administered PTSD Scale (CAPS) and the Patient Health Questionnaire 8 (PHQ8) interviews are gold standard questionnaires for the diagnosis of PTSD. Indeed, these questionnaires have several limitations and give an imprecise diagnosis. Indeed, self-report questionnaires used for mental disorder diagnosis have several limitations and biases, including (I) introspective ability: Participants may not be able to accurately self-assess their symptoms and experiences,
(II) rating scale bias: The use of simple yes/no responses or numerical scales can be limiting and subject to social desirability bias,
(III) memory biases: Participants may not accurately recall symptoms or experiences, leading to inaccurate self-reports,
(IV) language and cultural differences: Participants may have difficulty understanding and answering questions due to differences in language and culture,
(V) symptom over- or under-reporting: Participants may be more likely to either over-report or under-report their symptoms based on personal biases or the perceived social stigma associated with mental illness,
(VI) response bias: Participants may give socially desirable responses rather than truthful ones, leading to inaccurate self-reports.

Finding a technique that can overcome this issue, while being automated, would be welcome to help clinicians.
Numerous physiological changes, including heart rate, changes in blood pressure, changes in facial expression, and voice acoustics, can be used to indicate PTSD \cite{krothapalli2013characterization}. 
A technique to help clinicians and have more precise diagnostics is to use audio speech.
According to Scherer et al.\cite{scherer2015self}, psychomotor impairment, which affects articulation and motor control in PTSD patients, can be the cause of decreased vowel spacing. Three elements make up a standard speech-based PTSD recognition system: data collection, feature extraction, and classification.

\subsection{Deep Learning}
\label{sec:deep_learning}
Deep learning and processing power advancements now make it possible to measure speech-related indicators quickly and accurately. Deep learning is a potential method to close the gap between actual data and explanatory ideas in cognitive neuroscience and psychology. Deep neural networks are remarkably adept at learning to emulate human cognitive processes like facial recognition in a data-driven manner, but basic correlational techniques are restricted in their ability to distinguish useful relationships from spurious effects \cite{meehl1990summaries}, \cite{lecun2015deep}, \cite{lang1979bio}. 

Deep learning, however, is used to uncover stable probabilistic patterns in a data-driven manner, not to describe such theories precisely \cite{valiant1984theory}. Existing PTSD theories, like the Emotional Processing Theory  \cite{westra2007review}, are still extremely valuable and instructive for the selection of candidate predictors because the emphasis is on prediction rather than explanation \cite{shmueli2010explain}. 

The most advanced approach for describing sequential data currently is transformer networks \cite{vaswani2017attention}. Both (autoregressive) density modeling tasks and sequence-to-sequence modeling tasks fall under this category. Transformer networks use self-attention, a type of neural attention process \cite{luong2015effective}, to capture temporal dynamics in sequential data. Self-attention is unique in that it allows each position in a sequence to attend to every other position, making it possible to identify distant dependencies in the data. Additionally, it permits high-scale computation parallelization, which is not possible with earlier technologies. The attention computation is defined as :
\begin{equation}
    Attention(Q,K,V) = Softmax(Q.K^{T}).V
\end{equation}
$Q$ is the query, $K$ the key, and $V$ the value

Dense layers with ReLU activation functions are a common foundation for existing Transformer network formulations. However, a number of recent studies \cite{fang2022transformers} have demonstrated that adopting activation functions with some degree of stochasticity in their operation can result in significant performance gains, particularly in challenging machine learning tasks. \\

\section{Related work}

\subsection{Study of PTSD}
\label{study_ptsd}

The American Mental Association officially recognized post-traumatic stress disorder (PTSD) as a psychiatric disease in the third edition of the Diagnostic and Statistical Manual of Mental Disorders (DSM-III; American, 1980). The DSM-III and following DSM editions' definitions of PTSD are founded on the idea that, in contrast to other stressful experiences, traumatic events are etiologically connected to a particular syndrome \cite{DSMIII}. There are three symptom groups that make up the PTSD syndrome: (1) reliving the traumatic experience (1 out of 5 criterion symptoms is required), (2) avoiding stimuli that remind one of the incidents and becoming emotionally indifferent (3 out of 7 criterion symptoms are required), and (3) heightened arousal (2 out of 5 symptoms are required). The definition of these symptoms focuses on how they relate to the specific traumatic incident that has been recognized as the disorder's likely origin.
Recent reviews of PTSD empirical literature \cite{BUCKLEY20001041} found that PTSD subjects show a bias for trauma-related stimuli during the post-recognition. What comes out of it, is that PTSD patients are more susceptible to global valence effects at the initial stages of information processing. Additionally, there is some evidence to support the idea that autobiographical memory mechanisms in PTSD populations are comparable to those seen in depressive disorders. \\

\subsection{Machine Learning for PTSD detection with audio speech}
\label{ML_for_PTSD}

In the machine learning field, several approaches using audio speech data have shown promising results in PTSD detection. 
\newline In a study from 2019 \cite{banerjee2019deep}, a deep transfer approach has been proposed to improve PTSD diagnosis using audio. They created a deep belief network (DBN) model and transfer learning (TL) technique for PTSD diagnosis in their study. They calculated three different types of speech features and combined them using the DBN model. The TL approach was used from a speech recognition database, TIMIT, for PTSD identification. They demonstrated how the DBN performed better with the TL technique than the other common methods with an accuracy from 61.83\% to 74.99\%
\newline In \cite{gupta2022toxgb}, the authors suggested a method for PTSD detection that involves three steps: feature extraction, pre-processing or pre-emphasis, and classification. The pre-processing stage receives the incoming audio speech signal first, where the speech is divided into frames. The XGBoost-based Teamwork optimization (XGB-TWO) algorithm is then used to extract and classify the voice frame.
\newline Despite their very good performance, these approaches do not take advantage of time information.  \\

\subsection{Deep Learning for PTSD detection using audio speech}
\label{sec:DL_for_PTSD}
Deep learning approaches have also emerged. These approaches have the particularity of using one or multiple modalities.
\newline In some studies \cite{10.1145/3474085.3479236} \cite{josephine2022atypical}, researchers decided to detect PTSD through the perspective of emotions in the subject's speech. Both approaches use CNN-LSTM models. Their findings were on par with approaches that explicitly modeled the themes of the questions and replies, indicating that sadness may be identified through sequential modeling of an encounter with little knowledge of the interview's organizational framework, and is relevant in depression and PTSD detection. The performances of these approaches vary between 77\% and 98.68\%.
In the research from \cite{zhangPTSD2021}, the extraction of deep speech features is proposed using an autoencoder model based on bidirectional gated recurrent units (BiGRU), with the training target being the signal following cepstrum separation and the original speech serving as the network input. As input for the network in this model, they use the original speech, and the homomorphic speech serves as the model's training target. 
For the final decision or classification task, Support Vector Regression (SVR) was utilized after the model's long-term deep features and the Opensmile toolkit's short-term shallow features were passed to Random Forest (RF). This study uses the DAIC-WOZ data set to detect depression and reproduce the score from the PHQ-8 and PCL-C questionnaires. The RMSE (Root Mean Square Error) value is 5.68.

\section{Proposed approach}
\label{proposed_approach}

This study suggests the development and use of an innovative deep learning model for computer-aided detection of PTSD using audio recordings from clinical interviews intended to help with the identification of PTSD. The presented model is a cutting-edge transformer using stochastic layers, bottleneck attention, and stochastic depth, and only uses audio features.

The use of MFCC features is investigated in this study. The Mel-Frequency Cepstrum Coefficient (MFCC), used in the field of audio processing, is a representation of the short-term power spectrum of a sound that is based on a linear cosine transform of a log power spectrum on a nonlinear mel scale of frequency. As opposed to the linearly-spaced frequency bands used in the normal spectrum, the frequency bands used in the computation of MFCCs are equally spaced on the mel scale, which more accurately simulates the response of the human auditory system. This frequency warping allows better sound representation.
The equation \ref{eq:MFCC} shows how to compute MFCCs.
\begin{equation}
    \label{eq:MFCC}
    C_i = \sum_{n=1}^{N_f} Sn.cos[i.(n-0.5).\frac{\pi}{N_f}]; i\in [0;L]
\end{equation}
 $C_i$ is the $i$th MFCC coefficient, $N_f$ is the number of triangular filters used, $Sn$ is the log energy output of the nth filter, and L is the number of coefficients to compute.
\newline Additionally, MFCC processing can also help to remove noise and other artifacts from the audio signal, further improving the performance of deep learning models.
MFCC coefficients as low-level features that are fed to a deep learning model to learn high-level features for PTSD detection. The coefficients of each interview are registered in a matrix to be processed the same way images are processed. The embedding is similar to the one from ViTs. The matrix is transformed into a series of patches to be encoded. Once encoded, they fed the model.
\newline The combination of stochastic processes and MFCC can result in a better noise invariance, a more relevant information extraction \cite{jiang2021deep} and reduce overfitting.

\subsection{The model}
\label{sec:the_model}

\begin{figure*}[h!]
    \centering
    \includegraphics[width=.71\textwidth]{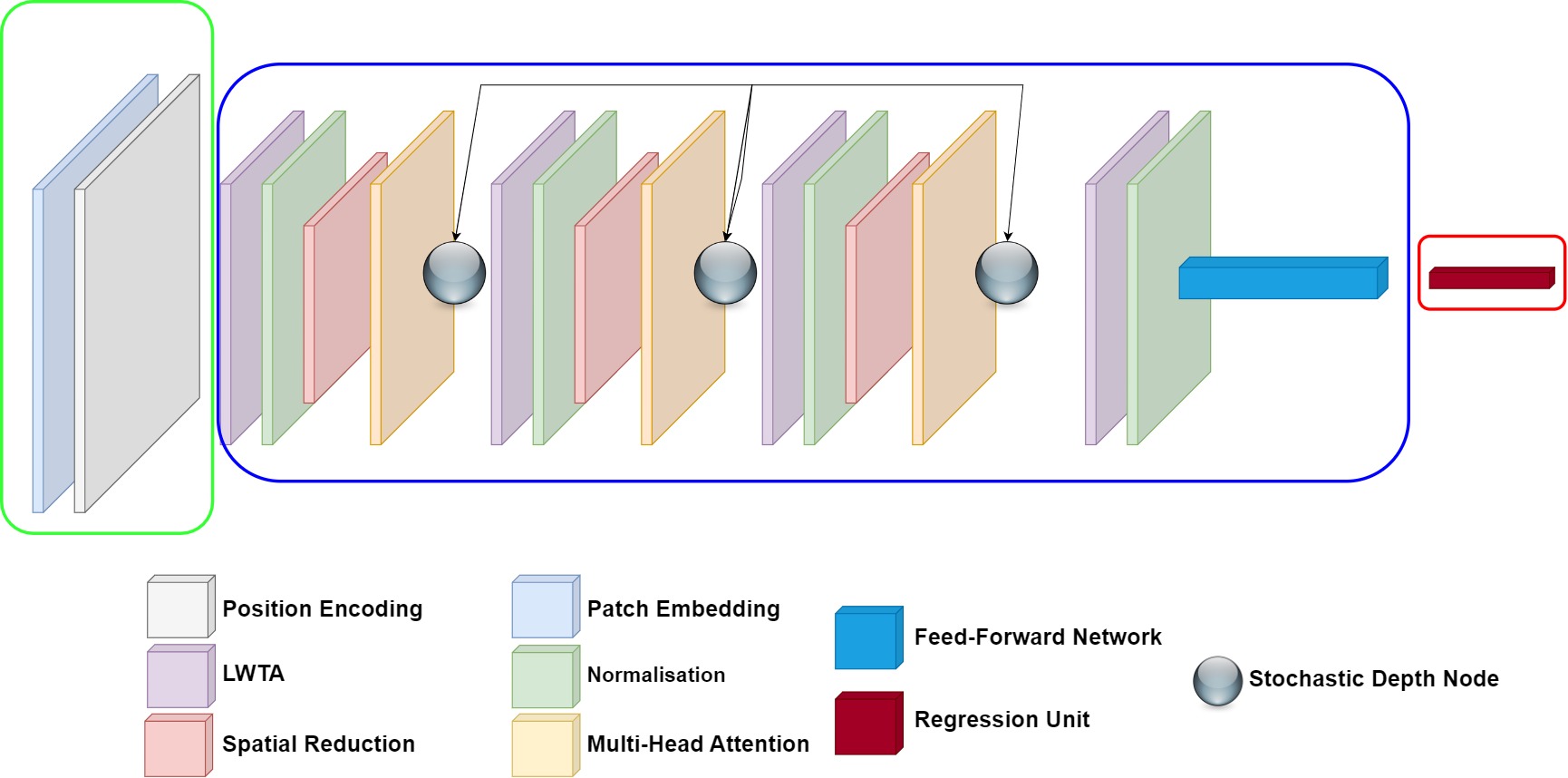}
    \caption{\textbf{Overall architecture of our Stochastic Transformer.} The model is composed of 3 main modules: the Patch Creation module (green frame), the Transformer Encoder module (blue frame), and the Regression module (red frame).}
    \label{fig:stochastic_transformer_architecture}
\end{figure*}

The proposed model is a stochastic transformer. This model's stochasticity consists of the use of layers operating stochastic operations (Local-Winner-Take-All (LWTA) mechanism and locally connected layers), stochastic activation functions, and in a stochastic depth, as shown in Figure.~\ref{fig:stochastic_transformer_architecture}.
The model is composed of a Patch Embedding layer that extracts patches from the input; a Position Encoding layer that encodes patches following the Fourier position encoding; an LWTA layer, described in \ref{sec:LWTA}; a Batch Normalization layer; a Spatial Reduction layer that reshapes the Attention tensor, to limit the amount of memory used; a Multi-Head Attention layer that performs self-attention; and a Stochastic Depth Node. The 4 last elements are repeated 3 times. Then, there is another LWTA layer; another Batch Normalization layer; and a feed-forward network. This feed-forward network is composed of dense layers and dropout layers, activated by the GeLU activation function. Finally, the regression unit is composed of a locally connected layer; an Average Pooling layer; a dropout layer; an LWTA layer, and a series of dense layers, activated by the GeLU activation functions.

\subsubsection{Local-Winner-Take-All layers}
\label{sec:LWTA}

In common dense layers, the dense layer is presented as a linear combination of the input vector $x \in \mathbb{R^J}$ obtained thanks to a weight matrix $W \in \mathbb{R^{J\times K}}$, to output a vector $y \in \mathbb{R^K}$.
LWTA layers are deep learning layers, where the units are not all connected to the other, and where only one unit, the one with the highest value, returns a non-null value.
In our implementation, an LWTA layer is a group of antagonistic linear units. The layer's input is initially sent to various blocks that engage in the competition via various neurons. Our method involves these linear units computing their activation within each block. Only one unit per block is selected. The units showing the maximal value are selected to create a mask indicating which elements in the tensor are equal to the maximum, as explained in the Equation.~\ref{eq:LWTA}, where $max(\cdot,\cdot)$ return the maximum between the 2 parameters. 

\begin{equation}
    \label{eq:LWTA}
    y = 
    \begin{cases}
        y_i; \quad if \quad y_i = max(0,x.W) \\
         0 \quad otherwise
    \end{cases}
\end{equation}

\subsubsection{Locally Connected layers}
\label{sec:LCN}
\begin{figure}
    \centering
    \includegraphics[width=0.4\columnwidth]{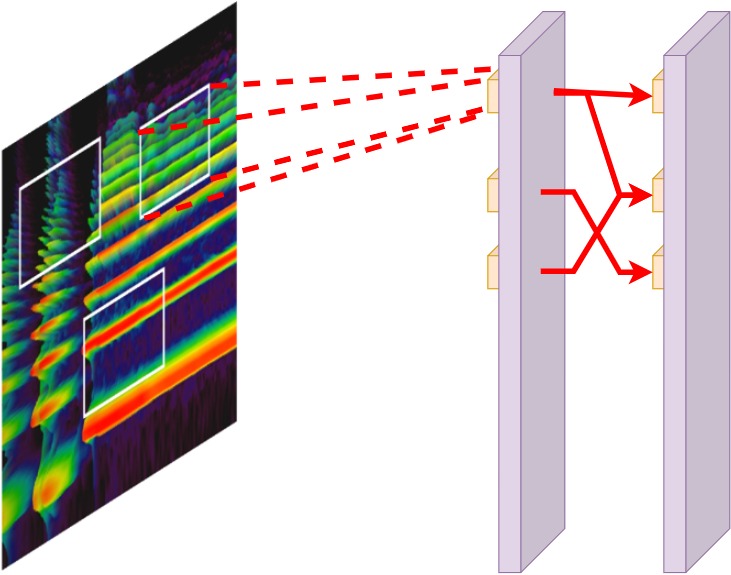}
    \caption{Scheme describing LCN. Neurons are not connected to all others neurons. LCN are also fed with a patch of the input, instead of the whole input, compared to fully connected layers.}
    \label{fig:lcn}
\end{figure}

To improve the regression unit of the model, the use of Locally Connected Layers (LCN) has been investigated. 
Unlike dense layers, units in LCN are not connected to all the others, as shown in Fig. \ref{fig:lcn}.
Local receptive fields, also known as patches, are the foundation of LCN and have a sparse but locally dense form. While concurrently limiting the complexity and latency of the model, LCN has the ability to compute discriminative features \cite{chen2015locally}. Patch computing is similar to convolution layers. In fact, they are almost the same with one subtle difference. Every output neuron in Convolutional layers uses the same filter (patches or pixels). The locally Connected Layer, on the other hand, includes individual filters for each neuron, that do not communicate with one another. 

\subsubsection{Dropout layer}
\label{sec:dropout}
Dropout is a regularization technique that randomly sets a fraction of input units to 0 during the forward pass of a neural network. This helps to prevent overfitting by making it difficult for any one neuron to rely too heavily on any particular input feature \cite{baldi2013understanding}. The stochastic process involved in Dropout is the random selection of which units to drop out during each forward pass. This selection is typically done independently for each unit, with a given probability. The units that are dropped out are not used in the computation of the output.

\subsubsection{GeLU activation function}
\label{sec:GeLU}
The Gaussian Error Linear Unit (GELU) activation function is a type of non-linear activation function that is used in deep learning \cite{hendrycks2016gaussian}. This activation function is defined as :

\begin{equation}
    \label{GELU}
    GELU(x) = x.\Phi(x) = x\frac{1}{2}[1 + erf(\frac{x}{\sqrt{2}})]
\end{equation}

$\Phi(x)$ is the Cumulative Distribution Function (CDF) of a standard normal distribution.
The stochasticity of the GELU activation function comes from the use of the CDF. It means that the output of GELU is a random variable, but the randomness is bounded by the cumulative distribution function of the standard normal distribution. Because of the use of the CDF function, the output of the GELU will always be between 0 and 1, which makes it a good choice as a probability output for some specific architecture like Transformers \cite{fang2022transformers}. This activation function also helps to improve the performance of the network by introducing a small amount of randomness in the forward pass.

\subsubsection{Stochastic Depth}
\label{sec:Stochastic_depth}
Stochastic Depth \cite{huang2016deep} is a process that overcomes the problem of vanishing gradients, overfitting, and long training time. Stochastic Depth is a training procedure that enables the seemingly contradictory setup to train short networks and use deep networks at test time. It starts with highly deep networks but drops a random selection of layers during training for each little batch before skipping over them with the identity function. The effectiveness of residual networks \cite{zagoruyko2016wide} and more modern Transformer with numerous transformer blocks \cite{zhang2021rest} are complemented by this strategy. Both the training time and test error are decreased.

\subsubsection{Spatial reduction of attention}
\label{sec:spat_reduc}
To limit the use of memory while computing attention, the Spatial Reduction Attention mechanism has been used, paired with Attention Bottleneck \cite{park2018bam}. With no reliance on domain-specific assumptions, this approach enables our architecture to use Transformers that are considerably deeper than those that are effective and use linear complexity layers, while limiting the use of Memory. By adding an asymmetry to the attention operation, our strategy directly applies attention to the inputs. Therefore, our attention compute becomes : 
\begin{equation}
    Attention = Softmax(\frac{Q.K^{T}} {\sqrt{d_{head}}}).V
    \label{eq:newAttention}
\end{equation}

\begin{equation}
    SR(x) = Norm(Reshape(x,Ri)W)
    \label{eq:spatial_reduction}
\end{equation}
Where $d_{head}$ is the dimension of the linear projection produced by the head of the transformer, $SR(\cdot)$ is the spatial reduction function used to reshape Attention, $Reshape(\cdot)$ is the reshape operation of the input,$Ri$ is the reduction ratio, and $W$ is a linear projection for the dimension reduction. Like in the previous definition, $Q$, $V$, and $K$ are respectively the query, the value, and key tensors.

\section{Experiments and Results}

\subsection{Dataset}
\label{sec:dataset}
The proposed model is trained and then tested on the Extended DAIC database (eDAIC) \cite{gratch2014distress}.
This dataset is made up of semi-clinical interviews intended to assist in the identification of mental health disorders like anxiety, depression, and post-traumatic stress disorder. These interviews were gathered as a component of a broader project to develop a computer agent that conducts interviews and recognizes verbal and nonverbal signs of mental diseases. The data includes 219 participant directories, each folder containing the raw audio data, eGeMAPS features, MFCC features, and Deep audio representations from VGG-16.
Each recording is labeled by the PHQ-8 and PCL-C scores. The PHQ-8 score determines the subject's level of depression severity, whereas the PCL-C determines their level of PTSD severity.

\subsection{Experimental setup}
\label{sec:exp_setup}
The model has been trained on an NVIDIA DGX Station. 2 GPUs have been used, and each of them has 32 GB of VRAM.
The optimizer and the hyperparameters have been chosen thanks to a hyperparameter optimization framework. The optimized parameters returned are the Adam optimizer with an initial learning rate of $10^{-3}$ and a weight decay of $10^{-4}$. 8 Transformer heads are used, as well as 3 Transformer blocks. The dropout rate is 0.2, and the survival probability for the stochastic depth is 0.2 too.
The metrics used to measure the performances of the model are the Root Mean Square Error (RMSE) and the CCC defined respectively by equation \ref{eq:RMSE} and equation \ref{eq:CCC}, where $K$ is the length of the output vector, $\hat{y_i}$ the $ith$ predicted output, $y_i$ is the $ith$ target, $COV(\cdot,\cdot)$ the mathematical covariance, $Y$ the output to predict, $\hat{Y}$ the predicted output, $\mu_Y$ is the mean of $Y$, $\mu_{\hat{Y}}$ is the mean of $\hat{Y}$, $\sigma_Y$ is the standard deviation of $Y$, and $\sigma_{\hat{Y}}$ is the standard deviation of $\hat{Y}$.

\begin{equation}
    \label{eq:RMSE}
     RMSE = \sqrt{\frac{1}{K}\Sigma_{i=1}^{K}{\Big(\hat{y_i}-y_i}\Big)^2}
\end{equation}

\begin{equation}
    \label{eq:CCC}
    CCC = \frac{2COV(Y,\hat{Y})}{(\mu_Y - \mu_{\hat{Y}})^2 + \sigma_Y ^2 + \sigma_{\hat{Y}} ^2}
\end{equation}


\subsection{Performances of our proposed approach for PTSD detection using audio}
\label{sec:perf_our}

In our experiments,  our model is trained to predict the PCL-C score of a subject by performing a regression using audio speech data. Our model shows an RMSE of 2.92 and a CCC of 0.533. The use of the attention mechanism allows us to benefit from temporal information, which is highly beneficial to process audio data. The use of the stochastic depth mechanism has been conducive to the obtained performances. Adding randomness by skipping multiple layers allows having a more robust model. LCNs are also efficient with the process of local information. Multiple filters can compute different discriminative features.

\subsection{Comparison with other methods}
\label{sec:comparison}
The Stochastic Transformer is compared with other models that were trained and tested on the eDAIC dataset with a different approach. We implemented and tested this approach.

\begin{table}[h!]
    \centering
    \resizebox{0.7\columnwidth}{!}{
    \begin{tabular}{|c||c|c|c|}
        \hline
        \multicolumn{4}{|c|}{Experimental results on the eDAIC dataset} \\
        \hline
        Model & Modality & RMSE & CCC \\
        \hline \hline
        KELM \cite{kaya2019predicting} & Text + audio & -- & 0.526 \\
        \hline
        GCNN-LSTM + CNN-LSTM\cite{rodrigues2019multimodal} & Text + audio & 6.11 & 0.403\\
        \hline
        Stochastic Transformer (our) & Audio & \textbf{2.92} & \textbf{0.533} \\
        \hline
    \end{tabular}
    }
    \caption{Table showing the performance of different models that has been trained on the eDAIC dataset. Models and approaches are different.}
    \label{tab:eDAIC_experiment}
\end{table}
In the first approach \cite{kaya2019predicting}, the authors used a Kernel Extreme Learning Machine (KELM). The features they used are the ones extracted thanks to an automatic speech recognizer. The primary idea behind this method is to create the first layer weights at random before learning the second layer weights by least squares regression.
\newline In\cite{rodrigues2019multimodal}, the authors investigated the use of Gated Convectional Neural Networks (GCNN) combined with LSTM and CNN combined with LSTM. The GCNN-LSTM model process textual features. The authors postulated that language characteristics would offer insightful information regarding the subject's mental health issues. Their model represents textual data by analyzing embedding in a bidirectional manner. The CNN-LSTM model process 1D audio.


Our Stochastic Transformers outperforms both approaches that were the most effective for PTSD diagnosis during the AVEC2019 challenge. 
The Stochastic Transformer shows an RMSE of 2.92 and a CCC of 0.533, compared to \cite{rodrigues2019multimodal}, where their approach shows an RMSE of 6.11, and a CCC of 0.403. Our approach is up to 52.21\% more effective. The KELM \cite{kaya2019predicting} shows a CCC of 0.526, which is 1.3\% lower.
\newline The capacity of Transformers to use temporal information allows them to have better performances \cite{liu2022tct}. Transformers do not rely on past hidden states to capture dependencies with previous patches. Due to this, there is no possibility of losing historical data. The link between several patches is also revealed through positional embedding and multi-head attention. The use of frequency-related features also results in good performances. MFCC contains more information compared to a 1D audio signal. These results can be explained by the use of stochastic depth. With this mechanism, the model inserts randomness during the process by skipping layers, which allows more robustness and better performance.

\subsection{Ablation study}
\label{sec:ablation_study}
To observe the effectiveness of stochastic units, an ablation study comparing different Transformers is conducted on the proposed model.

\begin{table}[h!]
    \centering
    \resizebox{0.7\columnwidth}{!}{
    \begin{tabular}{|c||c|}
        \hline
        \multicolumn{2}{|c|}{Results from different Transformer on the eDAIC dataset} \\
        \hline
        Model & RMSE \\
        \hline \hline
        Simple Transformer & 4.72 \\
        \hline
        Perceiver \cite{jaegle2021perceiver} (2 iterations) & 3.92\\
        \hline
        ST Encoder-Decoder \cite{voskou2021stochastic} & 3.72  \\
        \hline
        Stochastic Transformer(our) & \textbf{2.92} \\
        \hline
    \end{tabular}
    }
    \caption{Table showing the performance of different Transformers.}
    \label{tab:abalation study}
\end{table}

The ablation study shows that Transformers also outperforms other approaches, while the most preferment is our Stochastic Transformer showing an RMSE of 2.92, followed by another Stochastic Transformer from \cite{voskou2021stochastic}. The difference between both models is that a dedicated regression unit is composed of LCN, LWTA layers, and dropout layers. Another point to consider is that since there are 3 transformer blocks in the model architecture, it, therefore, has 3 MultiHeadAttention blocks. By using multiple MultiHeadAttention blocks, the model can attend to different, potentially conflicting sources of information in parallel, and then use the information from all heads to make a prediction. This allows the model to capture more complex relationships between input elements and can result in improved performance.
\newline Another reason why stochastic transformers can have better performance is that they can help the model explore a wider range of solutions during training. This is particularly important for transformer-based models, which have a large number of parameters and can easily get stuck in local optima. By introducing randomness during training, stochastic transformers can help the model explore different parts of the parameter space and find better solutions.


\section{Conclusion and future work}

This paper proposes an approach to detect PTSD thanks to a Stochastic Transformer. This approach uses audio data from a clinical interview. MFCC was extracted as a low-level feature to feed the neural network. Our approach shows better results on the eDAIC dataset compared to other methods thanks to the efficiency of Transformers on sequential data. The proposed method uses LWTA layers, locally connected layers, the GeLU activation function, and dropout layers. All these elements contain and exploit their stochastic properties, and allow us to obtain better performances. This approach allows better performances up to 52\%, compared to other methods on the same dataset. This article also shows that Transformers-based neural networks are more effective for the detection of PTSD if audio data is investigated, compared to the more common deep learning approach, thanks to the ability of transformers to establish relationships in the input sequence. For our future work, we will investigate the use of a multi-modal audio-visual and textual approach for computer-aided diagnosis of PTSD.


\section*{Acknowledgement}

Research financially supported by the French Ministry of Defence - Agence de l’Innovation de Défense (AID) under the number 94 - Thèses IDEES.

\bibliographystyle{splncs04}

\bibliography{references}
\end{document}